\documentclass[sn-mathphys,Numbered]{sn-jnl}
\usepackage{url}
\usepackage{graphicx}%
\usepackage{multirow}%
\usepackage{amsmath,amssymb,amsfonts}%
\usepackage{amsthm}%
\usepackage{mathrsfs}%
\usepackage[title]{appendix}%
\usepackage{xcolor}%
\usepackage{textcomp}%
\usepackage{manyfoot}%
\usepackage{booktabs}%
\usepackage{algorithm}%
\usepackage{algorithmicx}%
\usepackage{algpseudocode}%
\usepackage{listings}%
\begin{document}

\title[Article Title]{AI-GOMS: Large AI-Driven Global Ocean Modeling System}

\author[1]{\fnm{Wei} \sur{Xiong}}\email{xiongw21@mails.tsinghua.edu.cn}
\equalcont{These authors contributed equally to this work.}

\author[1]{\fnm{Yanfei} \sur{Xiang}}\email{xiangyf22@mails.tsinghua.edu.cn}
\equalcont{These authors contributed equally to this work.}

\author[2]{\fnm{Hao} \sur{Wu}}\email{wuhao2022@mail.ustc.edu.cn}

\author[1]{\fnm{Shuyi} \sur{Zhou}}\email{zhousy22@mails.tsinghua.edu.cn}

\author[1]{\fnm{Yuze} \sur{Sun}}\email{syz23@mails.tsinghua.edu.cn}

\author[1]{\fnm{Muyuan} \sur{Ma}}\email{mamy22@mails.tsinghua.edu.cn}

\author*[1]{\fnm{Xiaomeng} \sur{Huang}}\email{hxm@tsinghua.edu.cn}

\affil*[1]{\orgdiv{Department of Earth System Science, Ministry of Education Key Laboratory for Earth System Modeling, Institute for Global Change Studies}, \orgname{Tsinghua University}, \orgaddress{\city{Beijing}, \postcode{100084}, \country{China}}}

\affil[2]{\orgdiv{School of Computer Science and Technology}, \orgname{University of Science and Technology of China}, \orgaddress{\city{Hefei}, \postcode{230026}, \country{China}}}

%%====================================%%
%%              Abstract              %%
%%====================================%%

\abstract{Ocean modeling is a powerful tool for simulating the physical, chemical, and biological processes of the ocean, which is the foundation for marine science research and operational oceanography. Modern numerical ocean modeling mainly consists of governing equations and numerical algorithms. Nonlinear instability, computational expense, low reusability efficiency and high coupling costs have gradually become the main bottlenecks for the further development of numerical ocean modeling. Recently, artificial intelligence-based modeling in scientific computing has shown revolutionary potential for digital twins and scientific simulations, but the bottlenecks of numerical ocean modeling have not been further solved. Here, we present AI-GOMS, a large AI-driven global ocean modeling system, for accurate and efficient global ocean daily prediction. AI-GOMS consists of a backbone model with the Fourier-based Masked Autoencoder structure for basic ocean variable prediction and lightweight fine-tuning models incorporating regional downscaling, wave decoding, and biochemistry coupling modules. AI-GOMS has achieved the best performance in 30 days of prediction for the global ocean basic variables with 15 depth layers at 1/4$^{\circ}$ spatial resolution. Beyond the good performance in statistical metrics, AI-GOMS realizes the simulation of mesoscale eddies in the Kuroshio region at 1/12$^{\circ}$ spatial resolution and ocean stratification in the tropical Pacific Ocean. AI-GOMS provides a new backbone-downstream paradigm for Earth system modeling, which makes the system transferable, scalable and reusable.}


\maketitle

\section{Main}\label{sec1}

Ocean modeling drives oceanography into the era of experimentability and predictability. Employing numerical algorithms to solve the governing equations, ocean modeling enables researchers to simulate and predict ocean dynamics and facilitate qualitative and quantitative studies of the ocean\cite{sonnewald2021bridging}. With the development of high-performance computing clusters, more complicated numerical algorithms, parameterization schemes and data assimilation methods have been deploying in ocean modeling and operational ocean forecasting systems, which gradually reach accurate global ocean simulation\cite{hurlburt2009eddy}. It seems that traditional numerical ocean modeling have reveal an empirical knowledge that accurate simulations means higher computational costs. However, it is estimated that 10 billion times bigger in storage and faster in calculation than the current supercomputer to solve the primitive equations directly without approximations and parameterizations for high-fidelity global ocean simulation\cite{fox2019challenges}. With the development of the observation system and reanalysis technology, a new paradigm to model the ocean gradually shows its potential. A quiet revolution of ocean modeling is coming.

Abundant and high-quality data drive the booming development of deep learning models, which have emerged as one of the most highly discussed topics in the world. Generative artificial intelligence models, including large language models, have started to exert a profound influence on our daily lives. In the field of Earth science, artificial intelligence-based modeling has been achieving step-wise achievements and even being gradually deployed in operational institutions\cite{reichstein2019deep,duben2021machine,aires2021statistical}. For medium-range weather prediction, artificial intelligence-based modeling\cite{pathak2022fourcastnet,bi2022pangu,lam2022graphcast,nguyen2023climax,chen2023fengwu} has demonstrated that deep learning models with large parameters can perform comparably to numerical forecasting systems beyond the 10-day forecast\cite{chen2023fengwu}. These studies prove that large deep learning models have great potential to learn and model the Earth system dynamics. Furthermore, they exhibit superior scalability with increasing available data compared to numerical modeling. With much lower computational and temporal budget of the inference process, larger ensemble members\cite{pathak2022fourcastnet}, enhanced forecast timeliness and lower barriers to digital simulations will drive more possibilities for the Earth science application.

\begin{figure}[t!]
    \centering
    \includegraphics[width=1\textwidth]{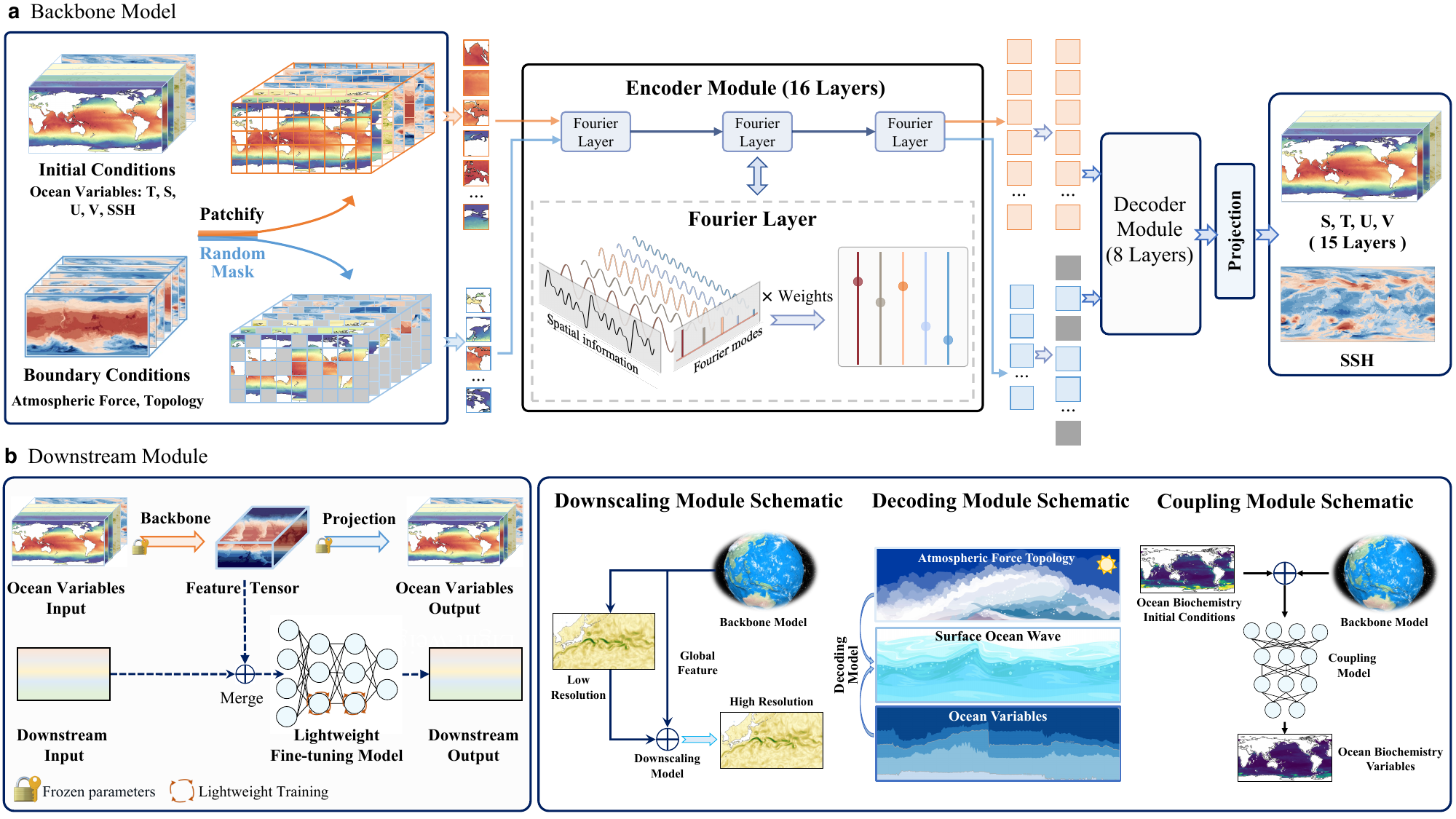}
    \caption{\textbf{Overall Architecture of AI-GOMS.} \textbf{a}, Backbone Model of AI-GOMS. Based on the asymmetric encoder-decoder structure with the random mask strategy, the model supports different length sequences. The prediction variables of the backbone model include 5 basic ocean variables (sea temperature(T), sea salinity(S), zonal velocity(U), meridional velocity(V) and sea surface height(SSH)) \textbf{b}, Downstream Module of AI-GOMS. Having trained the backbone model and frozen its parameters, lightweight fine-tune models can be applied to other ocean-related scenarios at a low cost.}
    \label{Fig1}
\end{figure}

However, it still has two gaps in artificial intelligence-based modeling in Earth system science. Firstly, the ocean, as one of the most important components of the Earth system, has not yet been modeled by the large deep learning model at present. Different from atmosphere simulation, the ocean is a special system with closed complex geometric boundaries, which means terrain topology constraints should be considered\cite{hurlburt2009eddy}. Ocean observations exhibit sparsity and an imbalanced distribution, and even large amounts are missing in the deep ocean\cite{levin2019global}. In addition, the eddy-resolving capability is a prerequisite for accurate global ocean simulation due to the magnitude of the Rossby deformation radius in the ocean, which requires finer resolution to resolve the eddies and their growth and decay in some regions\cite{roed1996modelling}. Secondly, existing studies have not fully explored the potential of the paradigm of large-scale deep learning models. These studies have not yet proven their model can handle different downstream applications well, which means their trained model has not been proven to be transferable to other related scenarios. Large models that are not reusable in various applications will not be able to take advantage of their low inference cost. Inspired by the gaps, we propose a large AI-driven global ocean modeling system named AI-GOMS with a general framework for three ocean-related downstream tasks. The overall architecture is shown in Fig.\ref{Fig1}.

AI-GOMS consists of a backbone model with the Fourier-based Masked Autoencoder structure and lightweight fine-tuning models incorporating regional downscaling, wave decoding, and biochemistry coupling modules. Inspired by the dynamic core of traditional ocean modeling, the backbone model of AI-GOMS is used to simulate and predict the basic global ocean physical variables (temperature, salinity, velocity and sea surface height). It is designed with an asymmetric encoder-decoder structure\cite{he2022masked} with Fourier-based attention blocks\cite{guibas2021adaptive}, which means the sequence lengths of the encoder and decoder can be various. Similar to the common vision transformer structure, our backbone model employs a patch embedding structure to map the various forms of physical fields into one-dimensional sequence tokens, which supports multimodal fusion\cite{dosovitskiy2020image,kim2021vilt}. Owning to the design, our model is available for two-dimensional and three-dimensional regular grid data as well as sparse data as input. Therefore, our model can be driven by multi-source data, which means it also natively supports data assimilation at the model level. Due to the asymmetric structure, we use a random mask strategy to selectively hide some patches of physics variables as input, which can be motivated to learn more intrinsic features rather than local spatio-temporal interpolation, to mitigate overfitting and get better performance in long-term prediction. For ocean-related downstream tasks, we exemplary present downstream modules to solve three different scenarios (regional downscaling, wave and biochemistry variables prediction). It proves that our trained backbone model can be transferred to other ocean-related scenarios by fine-tuning a simple module with extremely low training cost, which means users can download our trained backbone model to solve their own needs at a relatively low cost instead of training a whole model repetitively.

\section{Results}\label{sec2}
\subsection{Quantitative result and profile analysis on 30 days of prediction} \label{sec21}

We train our backbone model on the HYbrid Coordinate Ocean Model (HYCOM) global reanalysis data, which is the best estimation state for ocean variables with Global Ocean Data Assimilation Experiment (GODAE)\cite{chassignet2007hycom}. It is trained on 10 years of data (from 2000 to 2010), validated on 2011 and tested on 2012. Our model predicts five ocean variables daily, which include sea temperature, sea salinity, zonal and meridional stream velocity, with 15 depth layers (0m, 6m, 10m, 20m, 30m, 50m, 70m, 100m, 125m, 150m, 200m, 250m, 300m, 400m, 500m) and sea surface height at a spatial resolution of 1/4$^{\circ}$. Five atmospheric variables and topography data are introduced as boundary condition inputs to drive the model, which are from one of the best atmospheric reanalysis datasets named ERA5\cite{hersbach2020era5} and a topology dataset named ETOPO\cite{noaa2022etopo}. In 30 days scenario, our model shows strong stability and accuracy.

\begin{figure}[t!]
    \centering
    \includegraphics[width=1\textwidth]{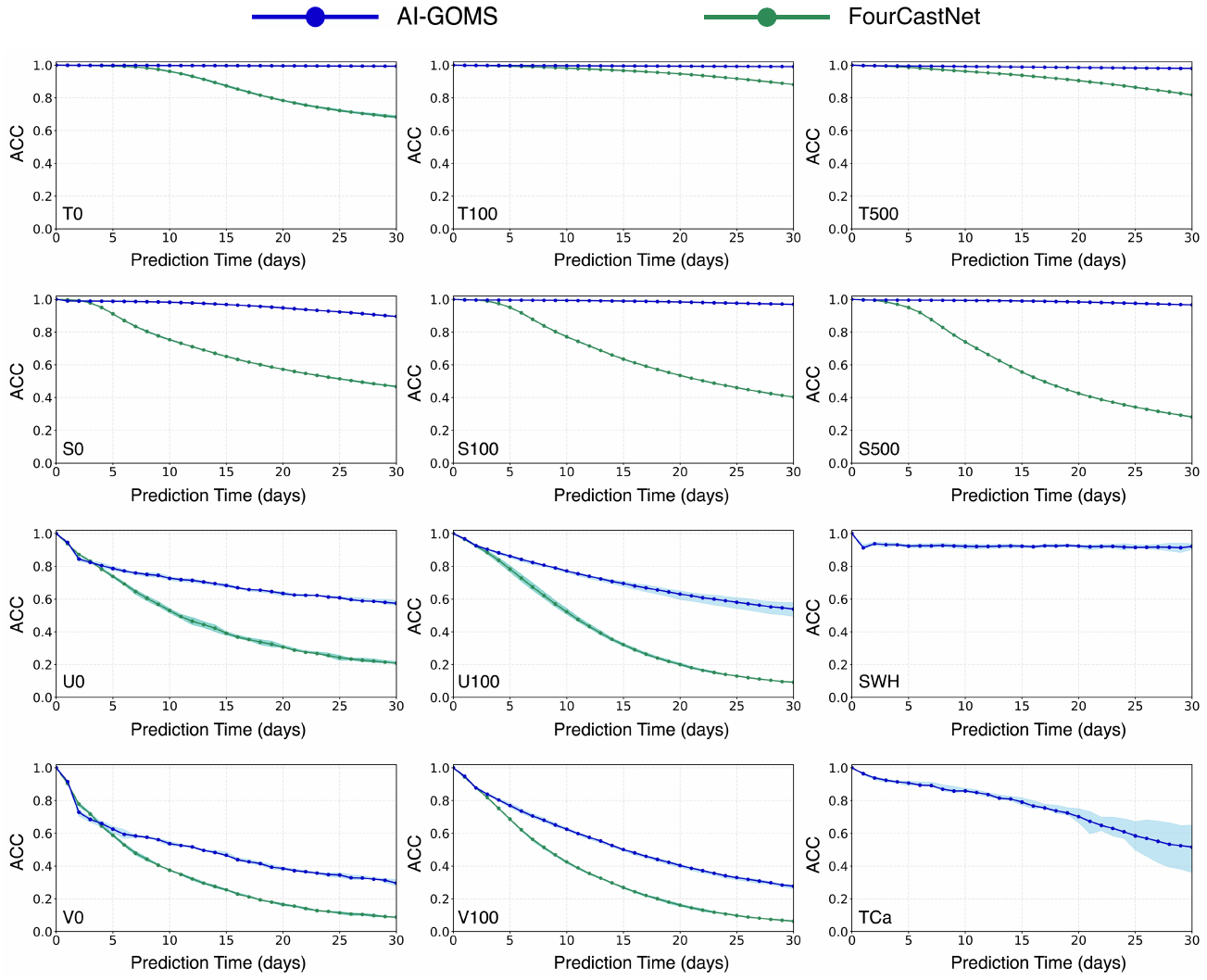}
    \caption{\textbf{AI-GOMS achieves the best performance for all variables in 30 days of prediction.} Latitude-weighted ACC for AI-GOMS (blue) and FourCastNet (green) averaged over several predictions in the out-of-sample testing dataset. The colored shaded regions around the ACC curves denote the region between the first and third quartile values. Here, T0, S0, U0 and V0 indicate the sea temperature, sea salinity, and the u-component and v-component of stream velocity at the sea surface, respectively. Similarly, T100, S100, U100, V100, T500 and S500 indicate that the variables are at 100 or 500 meters of depth. SWH and TCa indicate significant wave height and total chlorophyll a concentration.}
    \label{Fig2}
\end{figure}

In Fig. \ref{Fig2}, we measure the prediction accuracy of each ocean variable during testing using the latitude-weighted anomaly correlation coefficient (ACC) and compare AI-GOMS with FourCastNet\cite{pathak2022fourcastnet}, which is the state-of-the-art open source AI-based weather forecast model. AI-GOMS achieves better performance for all variables. Especially, AI-GOMS exhibits strong long-term predicting capability and has achieved significantly better performances in both ACC and RMSE (Root Mean Square Error).

\begin{figure}[t!]
    \centering
    \includegraphics[width=1\textwidth]{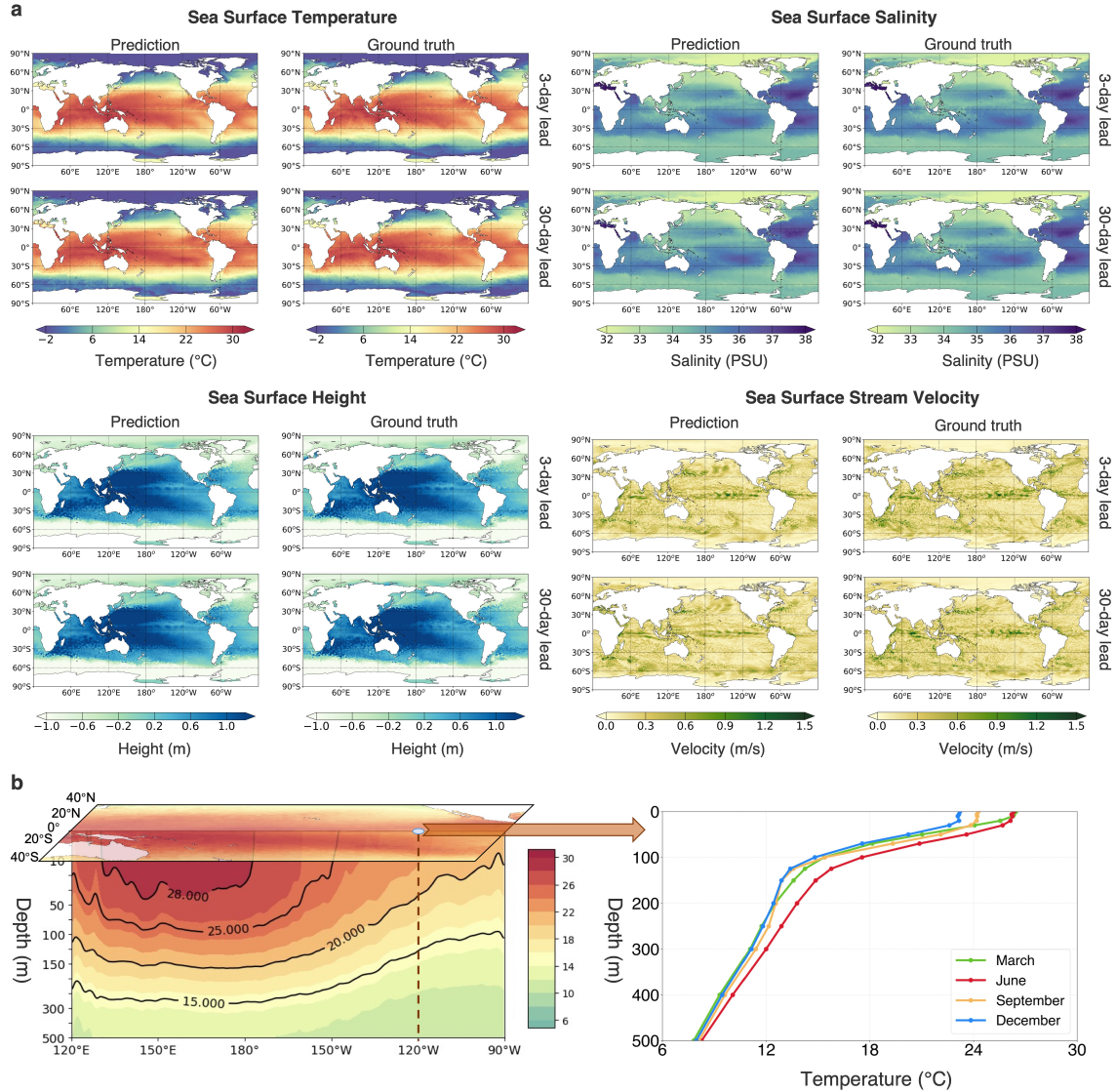}
    \caption{\textbf{Prediction Result of AI-GOMS Backbone Model.} \textbf{a}, The 3-day and 30-day predictions for main surface variables (sea surface temperature, sea surface salinity, sea surface height and sea surface velocity). For each case, the initial condition time is 12:00 UTC on 26 January 2012. \textbf{b}, The 30-day prediction sea temperature profile in the equatorial Pacific at 12:00 UTC on 26 December 2012 (left). The 30-day prediction time of sea temperature at the location (0$^{\circ}$,120$^{\circ}$E, right) is 12:00 UTC on 31 March 2012, 12:00 UTC on 29 June 2012, 12:00 UTC on 27 September 2012 and 12:00 UTC on 26 December 2012, respectively.}
    \label{Fig3}
\end{figure}

Beyond the better performance in statistical metrics, our model achieves good performance in the stratification of the ocean. The equatorial Pacific Ocean is an ideal region for profile analysis due to the presence of various equatorial waves, such as Rossby waves and Kelvin waves\cite{yanai1966stratospheric}. These movements deeply influence the feedback processes in air-sea interactions and climatic teleconnection. Therefore, we use the equatorial Pacific Ocean profile to evaluate our model. AI-GOMS accurately simulates equatorial sea temperature profiles that are influenced by the Walker Cell and Bjerkness feedback modulation\cite{bjerknes1969atmospheric}. Mixed layers and thermoclines can also be simulated, as shown in Fig.\ref{Fig3}. On the eastern coast of the Pacific Ocean, the rising thermocline and high chlorophyll concentrations due to upwelling can be simulated.

\subsection{Regional downsacling module for resolving mesoscale eddy}

For global ocean general circulation models, some special regions, such as Kuroshio and Gulf Stream, require finer spatial resolution because of their Rossby deformation radius\cite{roed1996modelling}. Accurate simulation of the pathway of Kuroshio and its extension requires the model to resolve mesoscale eddies and the fast processes in the ocean, which means finer spatial resolution and more stable algorithms\cite{yankovsky2022influences}. In other words, whether having the capability for eddy-resolving marks a watershed for ocean general circulation models.

Having considered that finer resolution is indispensable in such local regions, our architecture provides a regional downscaling module to support finer nested grids for the specific region. We tested the module in the region of Kuroshio and its extensions, downscaling the region from $1/4^{\circ}$ to $1/12^{\circ}$. As Fig.\ref{Fig4} shows, AI-GOMS can accurately simulate the stream pathway with a spatial resolution of $1/12^{\circ}$. The details of some mesoscale eddies are also accurately simulated. AI-GOMS can achieve an ACC of over 0.6 in 7 days of stream velocity prediction and over 0.6 in 7 days of sea surface height prediction in the Kuroshio region. According to the above results, the design of the downscaling module enables our system to achieve nested grid resolution to meet the diverse resolution requirements of various ocean regions.

\subsection{Wave decoding module for significant wave height prediction}

\begin{figure}[t!]
    \centering
    \includegraphics[width=0.8\textwidth]{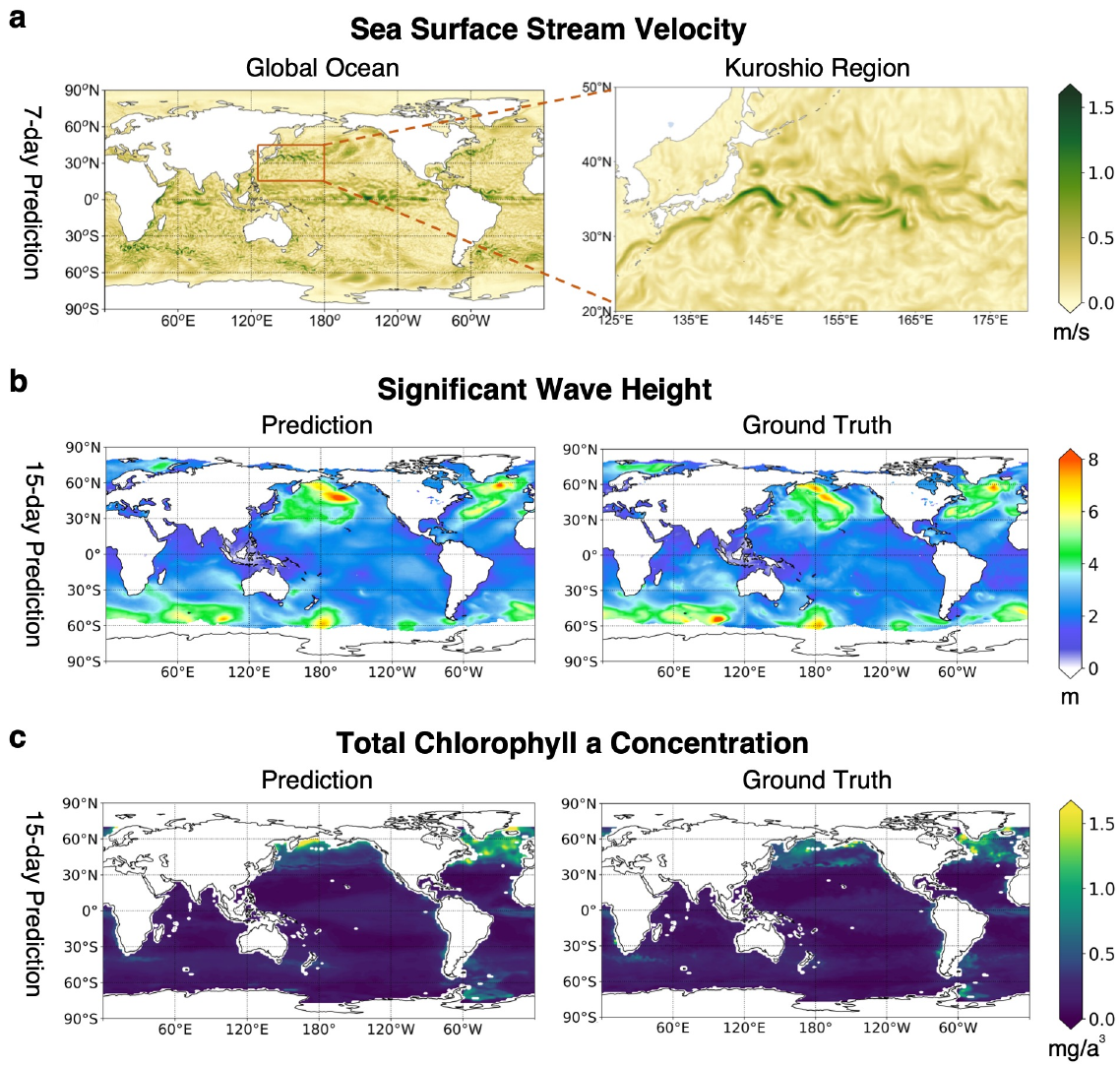}
    \caption{\textbf{Prediction Result of AI-GOMS Downstream Module.} \textbf{a}, The 7-day prediction of the Regional Downscaling Module for sea surface stream velocity at 12:00 UTC on 8 January 2012. \textbf{b}, The 15-day prediction of the Wave Decoding Module for significant wave height at 12:00 UTC on 16 January 2012. \textbf{c}, The 15-day prediction of the Biochemistry Couple Module for total chlorophyll a concentration at 12:00 UTC on 15 May 2012.}
    \label{Fig4}
\end{figure}

In operational oceanography, some important ocean prediction variables are derived from the basic ocean variables. Our architecture attempts to decode the operational prediction variables from the feature tensor and some closure conditions. Waves, driven by the wind, are classified into wind waves and swell, which are significant operational forecast variables. Therefore, we chose significant wave height as the decoding object variable in our experiments. Having lightweight fine-tuning trained, our architecture can realize the 30 days significant wave height prediction (Fig.\ref{Fig4}).

\subsection{Biochemistry coupling module for biochemical variables prediction}
From the perspective of marine science, the physical and biochemical properties of the ocean are highly correlated. In numerical ocean models, the physical ocean numerical model and the biochemical model are coupled using an external coupler to simulate biochemical variables\cite{fennel2022ocean}. Inspired by the design, we used a lightweight fine-tuning model to fuse the physical variables features and ocean biochemical conditions to predict 8 ocean biochemical variables (total chlorophyll a concentration, chlorophyte concentration, diatom concentration, coccolithophores concentration, cyanobacteria concentration, iron concentration, nitrate concentration and mixed layer depth). As Fig.\ref{Fig4} shows, our lightweight fine-tuning model gets good performance in 30 days of prediction. Our results show that the predictions of total chlorophyll a concentration exhibit high values along the eastern Pacific coastline, which corresponds to the upwelling predicted by the backbone model.

\section{Conclusion}\label{sec3}
In this paper, we design a large AI-driven global ocean modeling system named AI-GOMS for 30 days of simulation and prediction, which achieves global ocean simulation at 1/4$^{\circ}$ resolution with 15 layers. Our model not only achieves good performance in statistical metrics (RMSE and ACC), but also demonstrates good physical consistency in eddy simulations and vertical profiles.

In order to fully leverage the model's scalability and meet users' demands beyond the basic ocean variables, we design the downstream module for various ocean-related scenarios. We exemplary present three lightweight fine-tuning models (downscaling, decoding, and coupling) and achieve good performance at a lower cost. Our architecture explores the transferability capabilities of the trained backbone model.

Several future works are needed to further explore additional potential. First, our architecture supports sparse data as input, which will offer more possibilities for data assimilation and end-to-end training with sparse observation data. Second, more transfer learning techniques can be introduced into the architecture to further reduce fine-tuning costs.

\section{Method}\label{sec4}
\subsection{Architecture of AI-GOMS}
\subsubsection{Backbone Model}\label{sec411}
The overall structure of the AI-GOMS backbone is shown in Fig.\ref{Fig1}a, which is mainly established on the scheme of an asymmetric encoder-decoder and patch embedding-recovery module. Our model is designed to support multi-source data as input, including two-dimensional, three-dimensional and sparse data. However, all data needs to be preprocessed as grid-like data, which means sparse data should be processed into grid data with masks.

In our scenario, two-dimensional topology data, surface and near-surface variables ($1 \times 720 \times 1440$) and three-dimensional ocean variables with 15 layers($15 \times 720 \times 1440$) as boundary conditions and initial conditions. We stack these variables as the input tensor denoted as $X \in R^{C_{\text{in}} \times H \times W}$, where $C_{\text{in}}$ consists of different variables, and the size of the tensor is $67 \times 720 \times 1440$. We use the patch embedding module to non-overlapping patchify the input tensor and get the sequence of tokens $X \in R^{(h \times w) \times C_{\text{embed\_dim}}}$ (with a small patch size p × p, where p = 8). After the patch embedding module, part of the tokens will be truncated randomly by the specific mask strategy and the remaining tokens, along with a cosine positional encoding, will be encoded by several 1D-AFNO blocks. Subsequently, the encoded tokens, combined with their id-ranks, will be decoded and reconstructed by the decoder structure with shallower 1D-AFNO blocks. The decoded sequence of tokens will be mapped to the output variables $X_{\text{out}} \in R^{C_{\text{out}} \times H \times W}$ by the patch-recovery module with a two-dimensional transposed convolution operator.

More detailed explanations of our architecture, flow chart, additional visualizations of results and experimental logs are provided in Supplementary Information.

\subsubsection{Downstream Model}\label{sec412}

For various downstream tasks, our design adopts the backbone-downstream architecture. Having loaded the trained backbone model, the feature tensor of the backbone model is denoted by
$X_{f} \in \mathbb{R}^{C_{\text{decoder\_dim}} \times h \times w}$
is obtained from the input tensor
$X \in R^{C_{\text{in}} \times H \times W}$
calculated by the backbone model before the patch-recovery module. The input tensor of the downstream module is merged from the feature tensor of the backbone model $X_{f}$ and several conditions of the downstream scenario
$X_{d_{\text{in}}} \in R^{C_{d_{\text{in}}} \times H_{d} \times W_{d}}$, where $C_{d_{\text{in}}}$, $H_{d}$ and $W_{d}$ indicate the number of input conditions and the size of the longitude and latitude of the downstream scenario, respectively. We exemplary design the lightweight fine-tuning model for three downstream tasks. For the regional downscaling module, we test 8, 16 or 32 Residual Convolution blocks and a 3-fold upscaling ConvTranspose2d layer as the fine-tuning model. For the wave decoding module and biochemistry coupling module, we introduce two additional 1D-AFNO blocks and a projection layer as the fine-tuning model.

\subsection{Data and Training details}\label{sec42}

\subsubsection{Backbone Model}\label{sec421}
Our backbone model is trained by a publicly available and influential ocean reanalysis dataset from HYCOM, which mainly consists of 3-hourly of 8 sea surface variables and 9 ocean standard variables at a latitude and longitude resolution of $1/12^{\circ} \times 1/12^{\circ}$ from the sea surface to 5000m depth with 41 vertical layers from 1994 to 2015. The dataset is the optimal result from numerical simulation and multi-source observation (satellite and in-situ observation) by 3D-Var algorithm of Navy Coupled Ocean Data Assimilation (NCODA). In order to train a backbone model for daily prediction, we preprocessed the dataset and daily dataset at get $1/4^{\circ} \times 1/4^{\circ}$ spatial resolution, which contains enough information to resolve the majority of ocean regions. To accurately predict the thermocline and mixed layer of the ocean, we choose 15 depth layers (0m, 6m, 10m, 20m, 30m, 50m, 70m, 100m, 125m, 150m, 200m, 250m, 300m, 400m, 500m) of temperature, salinity, zonal stream velocity and meridional stream velocity.

\begin{table}[t]
\centering
\caption{Backbone Variables}\label{tab1}
\begin{tabular}{lll}
\toprule
Variable  & Layer & Details                                                              \\
\midrule
T         & 15    & Sea temperature ($^\circ$C)                         \\
S         & 15    & Sea Salintiy (PSU)                                                   \\
U         & 15    & Sea stream zonal velocity (ms$^{-1}$)                             \\
V         & 15    & Sea stream meridonal velocity (ms$^{-1}$)                         \\
SSH       & 1     & Sea surface height (m)                                               \\
U$_{10}$  & 1     & Wind meridonal velocity 10m from the surface (ms$^{-1}$)          \\
V$_{10}$  & 1     & Wind zonal velocity 10m from the surface (ms$^{-1}$)              \\
T2m       & 1     & Air temperature at 2m from the surface ($^\circ$C)  \\
MSL       & 1     & Mean surface level (m)                                               \\
SP        & 1     & Surface pressure (hPa)                                               \\
Topology  & 1     & Digital elevation and ocean bathymetry (m)  \\
\botrule
\end{tabular}
\end{table}

The goal of our backbone model is to predict the sea surface height (SSH), temperature, salinity, u-component (zonal) and v-component (meridional) stream velocity. In order to accurately learn complex dynamic relationships, we choose several boundary conditions, including the topology variable\cite{noaa2022etopo} and five atmospheric forcing variables\cite{hersbach2020era5}, as well as the initial conditions of sea surface height and four ocean basic variables, as input. Although the data contains 3D variables, such as ocean basic variables, as well as 2D variables, such as sea surface variables, each of the variables is represented as a 3D shape of C $\times$ 720 $\times$ 1440, where C represents the number of layers. We merge different variables by stacking them in the C dimension. Within each 24-hour period per day, we sample the variables from the dataset at 1200hours. We divide the dataset into training, validation and out-of-sample testing datasets, where the training dataset comprises the data from 2000 to 2010, the validation dataset is from 2011, and the testing dataset includes the data from 2012.

AI-GOMS employs auto-regressive prediction, which means it generates T-step prediction by iterating from the initial condition. The model predicts the later step by feeding its own output for the preceding time step combined with boundary conditions, similar to numerical ocean models. We define the input tensor as $X^{t}_{i}$ including 11 variables. These variables consist of two components: one is from the initial conditions or model output (ocean prediction variables), and the other is from boundary conditions (atmospheric forcing and topology variables). The output tensor, denoted as $X^{t}_{o}$, includes five variables. Here, $t$ represents the current time, and therefore, the next time step of the variables is represented as $X^{t+1}$, with a time interval of 1 day.

In the training process of the backbone model, we use the random mask strategy of gradually decreasing the mask ratio to force the model to learn hidden physics features behind sparse information. Two-step supervision is used in our training. We use the cosine learning rate schedule. The training process takes about 34.2 hours of wall-clock time on four clusters with 16 Nvidia A100 GPUs.

More details about data preprocessing, hyperparameters, training strategies and computational budget are provided in Supplementary Information.

\subsubsection{Downstream Module}\label{sec422}
\begin{table}[t]
\centering
\caption{Downstream Variables}\label{tab2}
\begin{tabular}{lllll}
\toprule
Variable & Resolution        & Details                                           & Task        & Source    \\
\midrule
SST      & 1/12$^{\circ}$    & Sea surface temperature ($^\circ$ C)              & Downscaling & HYCOM     \\
SSS      & 1/12$^{\circ}$    & Sea surface salintiy (PSU)                        & Downscaling & HYCOM     \\   
SSU      & 1/12$^{\circ}$    & Sea surface stream zonal velocity (ms$^{-1}$)     & Downscaling & HYCOM     \\   
SSV      & 1/12$^{\circ}$    & Sea stream meridonal velocity (ms$^{-1}$)         & Downscaling & HYCOM     \\   
SSH      & 1/12$^{\circ}$    & Sea surface height (m)                            & Downscaling & HYCOM     \\   
SWH      & 1/2$^{\circ}$     & Significant wave height (m)                       & Decoding    & ERA5      \\   
U$_{10}$ & 1/2$^{\circ}$     & 10m wind meridonal velocity (ms$^{-1}$)        & Decoding    & ERA5      \\  
V$_{10}$ & 1/2$^{\circ}$     & 10m wind zonal velocity (ms$^{-1}$)        & Decoding    & ERA5      \\
Tca      & 1$^{\circ}$       & Total chlorophyll a concentration (mg/m$^{3}$)    & Coupling    & NASA      \\   
Chl      & 1$^{\circ}$       & Chlorophyte concentration (mg/m$^{3}$)            & Coupling    & NASA      \\   
Dia      & 1$^{\circ}$       & Diatom concentration (mg/m$^{3}$)                 & Coupling    & NASA      \\   
Coc      & 1$^{\circ}$       & Coccolithophores concentration (mg/m$^{3}$)       & Coupling    & NASA      \\   
Cya      & 1$^{\circ}$       & Cyanobacteria concentration (mg/m$^{3}$)          & Coupling    & NASA      \\   
Irn      & 1$^{\circ}$       & Iron concentration (nano mole/L)                  & Coupling    & NASA      \\   
Nit      & 1$^{\circ}$       & Nitrate concentration (micro mole/L)              & Coupling    & NASA      \\   
MLD      & 1$^{\circ}$       & Mixed layer depth (m)                             & Coupling    & NASA      \\  
\botrule
\end{tabular}
\end{table}

The overall variables related to the three downstream tasks are shown in Table 2. The dividing method of training, validation and testing datasets is similar to that of the backbone model. The prediction time interval for all modules is 1 day.

\paragraph{Regional Downscaling Module}
We use the HYCOM dataset at a 1/12$^{\circ}$ resolution in the Kuroshio region, which ranges from 124.7$^{\circ}$ E to 180$^{\circ}$ and 21.28$^{\circ}$N to 45$^{\circ}$N as training label. We combined five low-resolution Kuroshio variables (sea surface temperature, sea surface salinity, zonal and meridional surface stream velocity and sea surface height) and the feature tensor of the backbone model with global information. The size of the condition tensor is 5 $\times$ 120 $\times$ 230 at 1/4$^{\circ}$ spatial resolution in time step $t$ and the feature tensor is 512 $\times$ 90 $\times$ 180. Having through the lightweight fine-tuning model, the size of the output tensor is 5 $\times$ 360 $\times$ 690 in the next time step $t+1$, which achieves 3-times super resolution from 1/4$^{\circ}$ to 1/12$^{\circ}$.

\paragraph{Wave Decoding Module}
We use significant height wave dataset as prediction variables in this scenario. Having considered the mechanics of ocean waves, we set the 10m wind velocity field from the surface as boundary conditions to drive the model. These variables are all from the ERA5 dataset\cite{hersbach2020era5}. In this scenario, the input variables beside the feature tensor are combined the initial condition of a significant height wave and the boundary condition of two components of the 10m wind velocity field from the surface. The size of the tensor is 3 $\times$ 360 $\times$ 720. Our lightweight fine-tuning model maps the wave variables into a next time step at each iteration. Besides, we use the annual mean sea ice distribution area as the mask of sea ice.

\paragraph{Biochemistry Coupling Module}
Our Biochemistry Coupling Module is trained by 8 biochemical variables, which are from the NASA Ocean Biochemical Model assimilating satellite chlorophyll data\cite{gregg2017nasa}. The dataset integrates satellite remote sensing observations and ocean circulation-biochemical coupled numerical models. We re-grid each of the variables into a global regular Euclidean grid at 180 $\times$ 360 by the bilinear interpolation method\cite{schulzweida_uwe_2022_7112925}. After the re-gridding process, we combine 8 biochemical variables (total chlorophyll a concentration, chlorophyte concentration, diatom concentration, coccolithophores concentration, cyanobacteria concentration, iron concentration, nitrate concentration and mixed layer depth) as initial conditions and feature tensors of the backbone model with physical oceanography information. Through the lightweight fine-tuning model, 8 biochemical variables in the next time step are predicted.

\subsection{Statistical Metric}\label{sec43}
Following the former works\cite{rasp2020weatherbench,pathak2022fourcastnet}, we use latitude-weighted RMSE and ACC as statistical metrics to assess the performance. Land area is not calculated for each metric. Therefore, we denote $(i,j)$ as the index of grid coordinates in the ocean grid set $G$ as follows.

\paragraph{Latitude Weighting}

The latitude weighting $L(j)$ for the latitude at the j-th index is defined as
$$
L(j)=\frac{\cos (\operatorname{lat}(j))}{\frac{1}{N_{\mathrm{lat}}} \sum_j^{N_{\mathrm{lat}}} \cos (\operatorname{lat}(j))},
$$
where $N_{lat}$ is the number of latitude index.

\paragraph{Root mean square error}

The root mean square error (RMSE) for the variable $c$ at forecast time step $t$ is defined as

$$
\operatorname{RMSE}(c, t)=\sqrt{\frac{1}{|G|} \sum_{i, j} L(i)\left(\mathbf{X}_{\mathrm{pred}}^{t}[c, i, j]-\mathbf{X}_{\mathrm{true}}^{t}[c, i, j]\right)^2},
$$
where $\mathbf{X}_{\mathrm{pred}}^t$ and $\mathbf{X}_{\mathrm{true}}^t$ represents the value of predicted and true variable $c$ at coordinates $(i,j)$.

\paragraph{Anomaly correlation coefficient}

The anomalous correlation coefficient (ACC) for the variable $c$ at the forecast time step $t$ is defined as

$$
\operatorname{ACC}(c, t)=\frac{\sum_{i, j} L(i) \tilde{\mathbf{X}}_{\mathrm{pred}}^{t}[c, i, j] \tilde{\mathbf{X}}_{\mathrm{true}}^{t}[c, i, j]}{\sqrt{\sum_{i, j} L(i)\left(\tilde{\mathbf{X}}_{\mathrm{pred}}^{t}[c, i, j]\right)^2 \sum_{i, j} L(i)\left(\tilde{\mathbf{X}}_{\mathrm{true}}^{t}[c, i, j]\right)^2}},
$$
where $\tilde{\mathbf{X}}_{\mathrm{pred}}^t$ represents the climatological-mean-subtracted value of the predicted and true variable $c$ at coordinates $(i,j)$. The climatological mean is calculated by sampling HYCOM reanalysis data at 2-year intervals between 2000 and 2010.

\backmatter

\clearpage

\bibliography{sn-bibliography}

\end{document}